\documentclass{webofc}
\usepackage[varg]{txfonts}   
\begin{document}
\title{Clustering of Hotspots in the Cosmic Microwave Background}
%
%

\author{\firstname{} \lastname{Low} \firstname{En Zuo Joel} \inst{1} \and
        \firstname{Abel} \lastname{Yang}\inst{1}\fnsep\thanks{\email{phyyja@nus.edu.sg}}
}

\institute{Department of Physics, National University of Singapore}

\abstract{%
The physics behind the origin and composition of the Cosmic Microwave Background (CMB) is a well-established topic in the field of Cosmology. Literature on CMB anisotropies reveal consistency with Gaussianity \cite{PlanckRes16}, but these were conducted on full multi-frequency temperature maps. In this thesis, we utilise clustering algorithms to specifically conduct statistical analyses on the distribution of hotspots in the CMB. We describe a series of data processing and clustering methodologies conducted, with results that conclusively show that the counts-in-cells distribution of hotspots in the CMB does not follow a Poisson distribution. Rather, the distribution exhibits a much closer fit to both the Negative Binomial Distribution (NBD) and the Gravitational Quasi-Equilibrium Distribution (GQED). From this result, we conclude that structure likely existed in the early universe, from the period of the recombination Epoch, possibly opening new insights in the field of galaxy formation.}
\maketitle

\section{Introduction}
\label{intro}
\subsection{Cosmic Microwave Background}
\label{sec-1-1}
In the current theory of Big Bang Cosmology, the cosmic microwave background (CMB) is leftover electromagnetic radiation from the primordial stages of the formation of our universe. While the CMB may be consistent with Gaussianity on a whole \cite{PlanckRes16}, detailed observations reveal pockets of anisotropy scattered across the distribution, forming a pattern similar to that of a hot gas that has expanded over time. Till date, no studies have been conducted specifically on the statistical nature of the distribution of hotspots in the CMB. As such, we utilise clustering algorithms to specifically conduct statistical analyses on the distribution of these hotspots.

\subsection{Hypothesis}
\label{sec-1-2}
Through clustering analysis of CMB data, we seek to compare the distribution of hotspots in the CMB with the negative binomial distribution (NBD) and gravitational quasi-equilibrium distribution (GQED). These distributions were chosen as they provide a good description of the counts-in-cells distributions of galaxies \cite{CountsObs,CountGQED}.

If the GQED or NBD in fact proves to be a close match to the distributions of hot spots in the CMB, we will be able to conclude that matter in the recombination epoch already possessed some form of structure.

\section{Materials \& Methods}
\label{Methods}
\subsection{Planck CMB Temperature Map}
\label{sec-2-1}
The Planck Space Telescope was sent into orbit in May 2009 by the European Space Agency (ESA) to survey the CMB. Out of the published CMB maps, the SMICA product is labelled as preferred by the ESA and as such, was selected for further analysis \cite{PlanckRes9}. A Heaviside filter (95th percentile) was applied on the CMB Temperature map. The end result is a list of boolean HEALpix map, from which we can obtain the galactic longitude and latitude of each hot pixel.

To exclude possible sources of foreground contamination from the galaxy, such as galactic dust that lie in the plane of the milky way, we apply a mask on the dataset by excluding all points $10^\circ$ above and below the galactic equator.

\subsection{Clustering Algorithm: HDBSCAN}
\label{sec-2-2}
Conventional clustering algorithms (e.g. $k$-means) make assumptions that do not necessarily hold true for the SMICA dataset. We expect the real data to be noisy which can serve create apparent bridges between two separate clusters, leading to inaccurate clustering results. We also expect clusters of varying densities, and non-spherical clusters to be a concern. The total number of hotspots in the dataset was also an unknown.

HDBSCAN is an acronym which stands for “Hierarchical Density-Based Spatial Clustering Application with Noise” and was eventually selected as the clustering algorithm of choice as it was able to succinctly deal with the above stated conditions \cite{HDBSCAN}.

\subsection{Probability Distributions }
\label{sec-2-3}
The galaxy counts-in-cells (CIC) distribution describes the spatial location of galaxies. It can be generalised to a form of $f(N,V)$, giving the probability of finding $N$ hot spots in a region of volume $V$, or solid angle $\Omega$. In the $f_V(N)$ form, the volume is taken to be constant.

The probability mass function and variance of the GQED is as follows:
\begin{equation}
f_V(N) = \frac{\bar{N}(1-b)}{N!}\left(\bar{N}(1-b)+Nb\right)^{N-1}e^{-\bar{N}(1-b))-Nb}
\end{equation}
\begin{equation}
\langle(\Delta N)^2\rangle = \frac{\bar{N}}{(1-b)^2}
\end{equation}
where $\bar{N}$ is the mean of the distribution and $b$ is a clustering parameter between 0 and 1.

The probability mass function and variance of the negative binomial distribution (NBD) is as follows:
\begin{equation}
f_V(N) = \frac{\Gamma\left(N+\frac{1}{g}\right)}{\Gamma\left(\frac{1}{g}\right)N!} \frac{\bar{N}^N\left(\frac{1}{g}\right)^{\frac{1}{g}}}{\left(\bar{N}+\frac{1}{g}\right)^{N+\frac{1}{g}}}
\end{equation}
\begin{equation}
\langle(\Delta N)^2\rangle = \bar{N}^2 g +\bar{N}
\end{equation}
where $\bar{N}$ is the mean of the distribution and $g$ is a parameter greater than 0 related to the variance.

\subsection{Steps taken to obtain CIC distribution of hotspots in CMB}
\label{sec-2-4}
From the masked and filtered CMB data, we use HDBSCAN to obtain the positions of cluster centres, using a haversine metric on the galactic longitude and latitude of each hot pixel.

To lay down the cells, we spread cell centres evenly across the unmasked area. The physical extent of each cell is the cap with radius $\theta$ on the surface of a sphere goving a solid angle of $4\pi \sin^2(\theta/2)$. We then count the number of hotspots that lie within each cell to obtain the counts-in-cells distribution.
%

\section{Results}
\label{Results}
\begin{figure*}
\centering
\includegraphics[width=0.48\textwidth,clip]{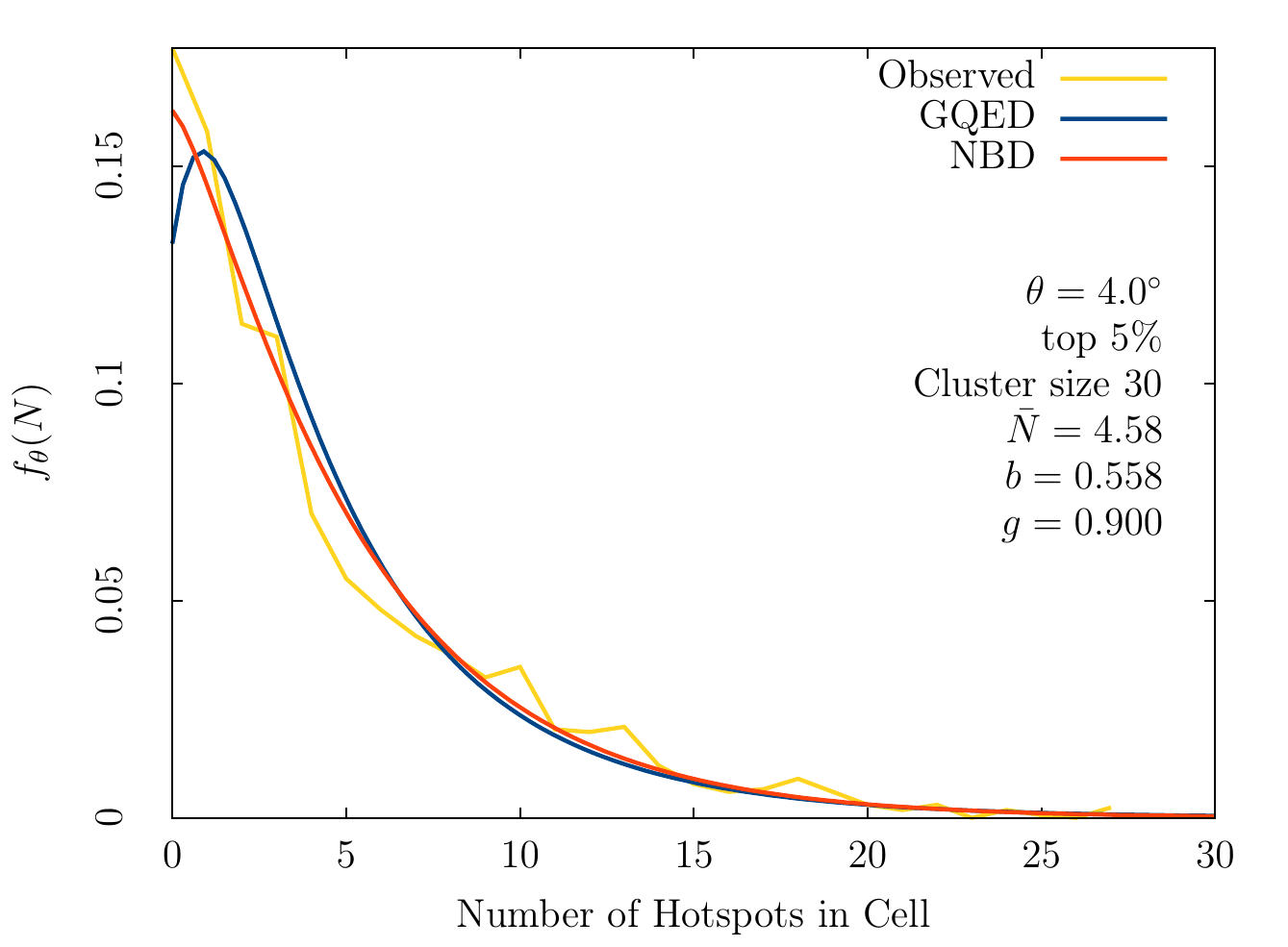}
\includegraphics[width=0.48\textwidth,clip]{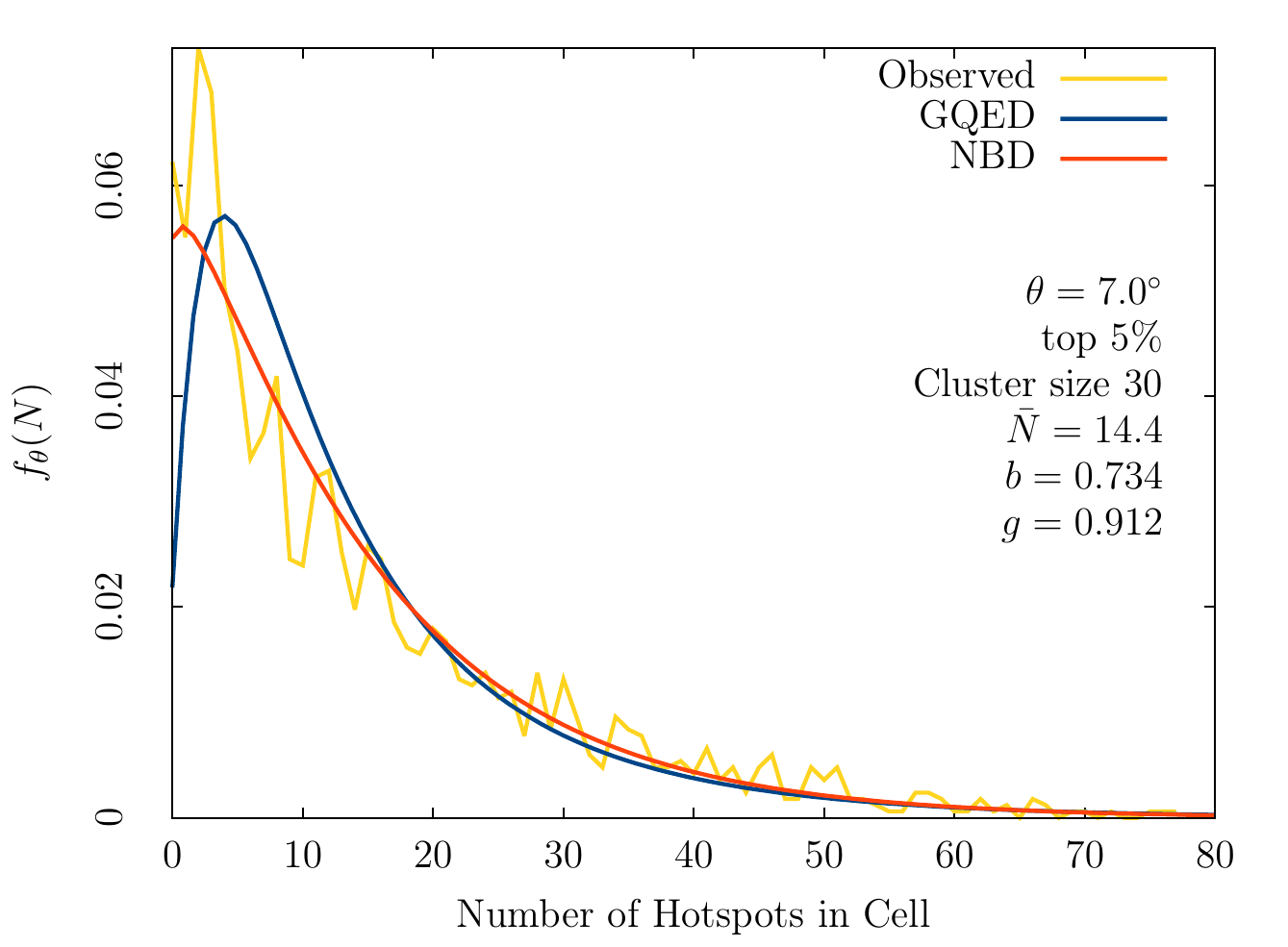}
\caption{Comparison between NBD and GQED for cell sizes of $4.0^\circ$ and $7.0^\circ$}
\label{fig-2}       
\end{figure*}

\subsection{Least squares goodness of fit measure}
\label{sec-3-1}
The least squares goodness of fit (LSGF) measure presents a quantitative measure of the goodness of fit between an observed, and expected distribution, given by
\begin{equation}
X = \sum_{i=0}^N \left(O_i-E_i\right)^2.
\end{equation}
This is the sum of the squares of the difference between the expected and observed $f_V(N)$. Smaller values of $X$ indicate a closer fit.

From figure \ref{fig-2}, we see the NBD consistently outperforms the GQED, with a smaller LSGF value across most of the tested parameter sets.

\subsection{Accounting for variance by resampling}
\label{sec-3-2}
Variations in the distribution of hotspots in the CMB can result in sub-volumes which are not statistically similar. To obtain an indication of the possible variations across the sky, we conduct a resampling procedure where for each instance, we exclude 25\% of the cells based on the galactic longitude of the cell centre. We obtain separate instances by shifting the window by $\pi/3$ radians each time, with an overlap of $\pi/6$ radians.

The minimum and maximum histogram values are then combined to obtain a window of confidence. An optimum result would be a tight window with only one of the fitted probability distribution functions falling into it, indicating an accurate fit with high confidence.

\begin{figure*}
\centering
\includegraphics[width=0.48\textwidth,clip]{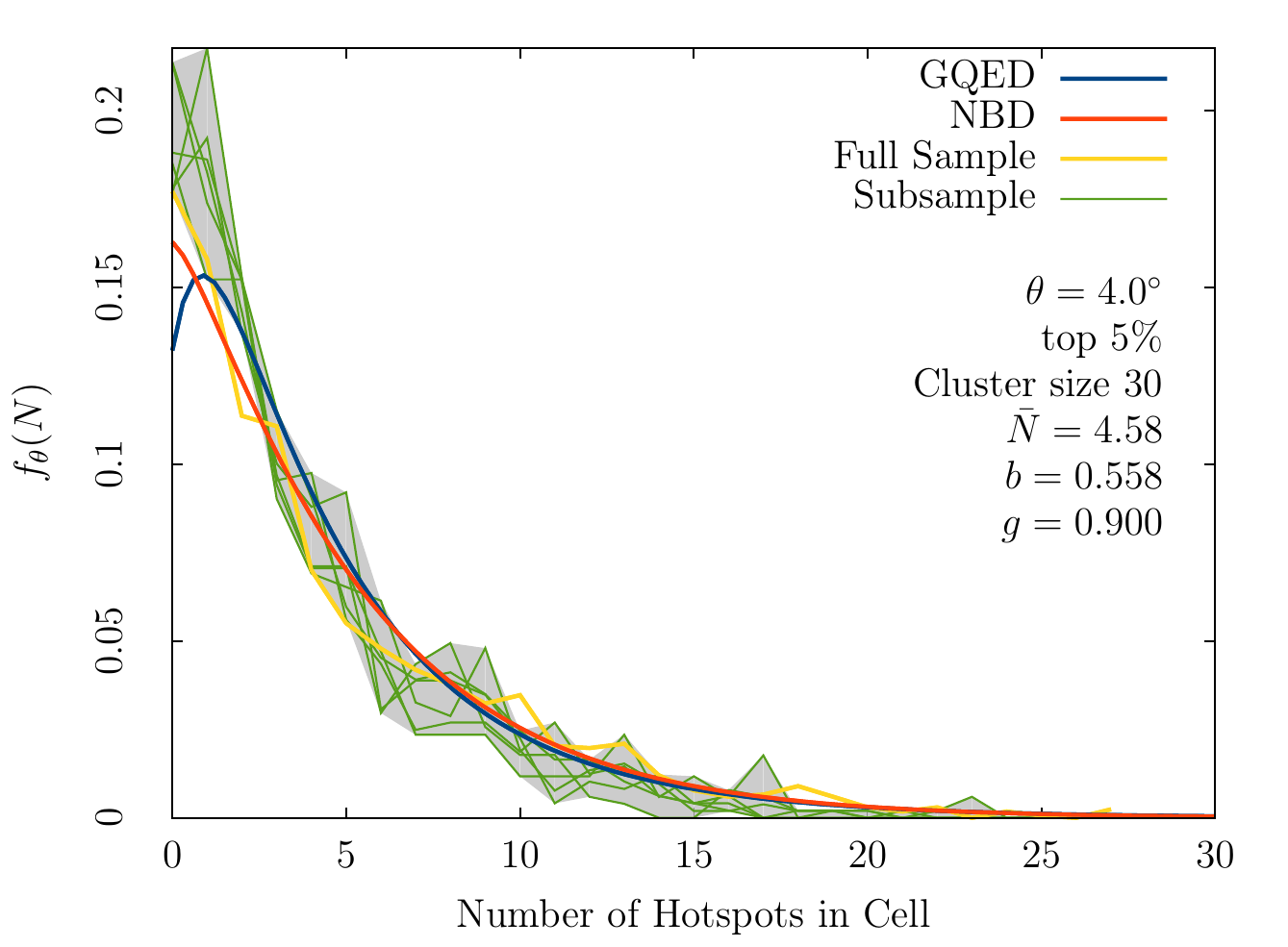}
\includegraphics[width=0.48\textwidth,clip]{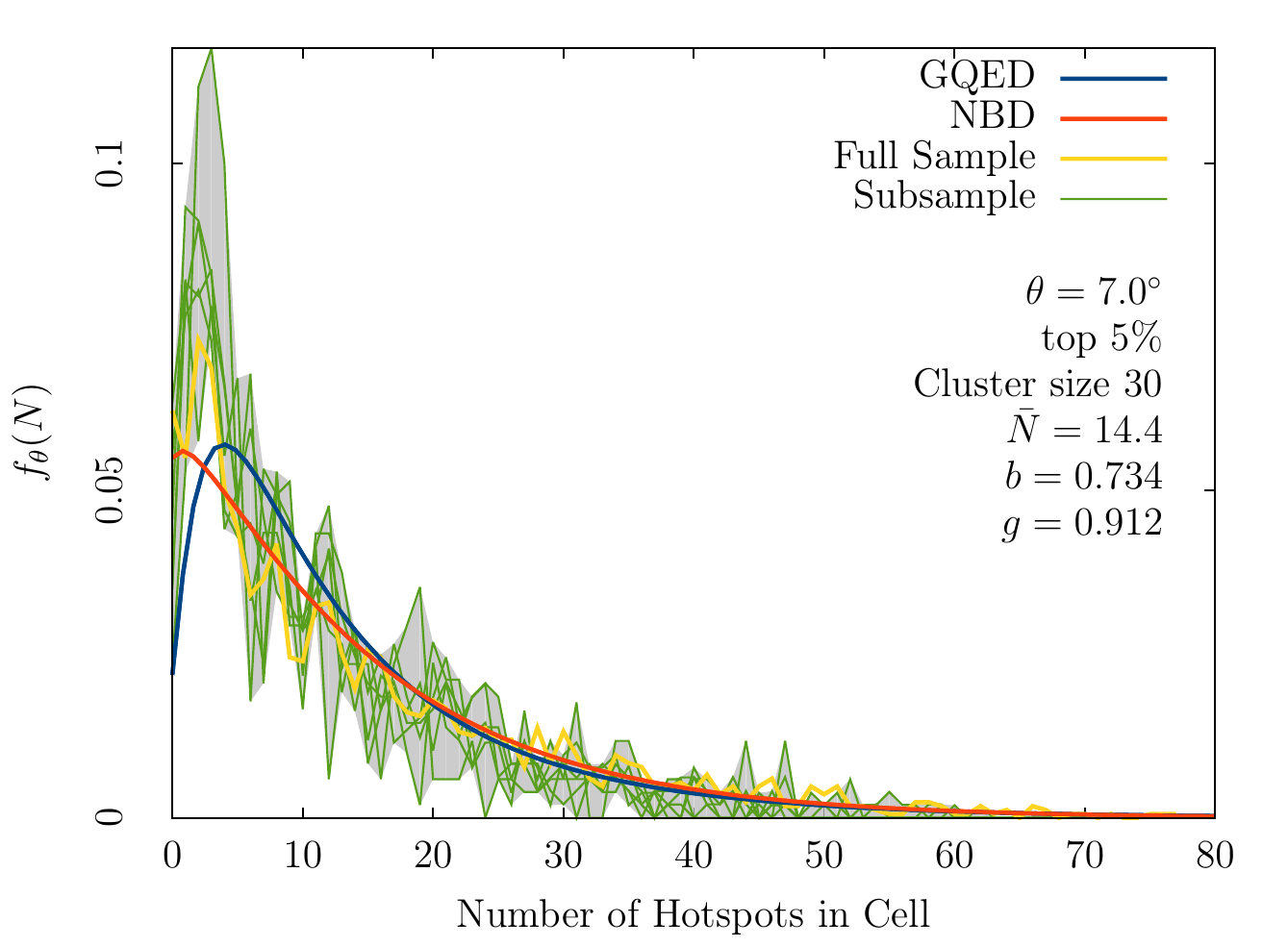}
\caption{Counts-in-cells for different resampling windows for cell sizes of $4.0^\circ$ and $7.0^\circ$}
\label{fig-3}       
\end{figure*}

From figure \ref{fig-3}, we observe that the NBD and GQED trace out almost identical paths, diverging only at low values of $N$. There are however large variations close to small values of $N$, suggesting the strong presence of noise in the data. This noise is likely to have originated from cosmic variance in the data collected by ESA, and it explains the greater impact in the small $N$ region.

\section{Discussion \& Conclusion}
\label{Conclusion}
After obtaining all LSGF values, we compared the results between the NBD and GQED. Out of all 396 parameter choices, the counts-in-cells distribution follows the NBD and GQED somewhat closely, indicating that the hotspots in the CMB are in fact not randomly distributed.

Comparing the closeness of fit between the NBD and GQED, we observe that the GQED generally under predicts the number of voids where $N = 0$. This leads to the NBD presenting a closer fit at small values of $N$. However, we are unable to conclusively state that the NBD is a closer fit as the resampling window is noisy near low $N$ regions. Also, the NBD is unphysical description of galaxy clustering, violating the 2nd law of thermodynamics \cite{ThermoRes}.

For future work, we intend to use a more comprehensive galactic foreground contamination mask and higher resolution temperature maps. We also intend to consider different percentile cut-offs for the initial Heaviside filter (e.g. 1-$\sigma$ instead of just 95th percentile).

\end{document}